\documentclass[letter]{aa} 

\usepackage[breaklinks,pdfnewwindow=true,colorlinks=true,citecolor=blue]{hyperref}

\usepackage{graphicx}
\usepackage{linenoaa}
\usepackage{txfonts}
\usepackage{rotating}
\usepackage{amsmath}
\usepackage{longtable}
\usepackage{tablefootnote}
\usepackage{subcaption}
\usepackage{dblfloatfix}
\usepackage{float}
\usepackage{amssymb}
\usepackage{xcolor}
\usepackage{booktabs}
\usepackage[version=3]{mhchem}
\usepackage{mathrsfs}
\usepackage{multirow,bigdelim,dcolumn,booktabs}
\usepackage{mathabx}
\usepackage{graphicx,caption,subcaption}
\usepackage{orcidlink}
\newcommand{\orcid}[1]{\unskip\protect\href{https://orcid.org/#1}{\protect\includegraphics[width=8pt,clip]{logo_orcid}}}
\newif\ifAMStwofonts

\newcommand{\Msun}{\rm M_{\odot}}
\newcommand{\Lsun}{\rm L_{\odot}}

\newcommand{\IRSFAUST}  {L1551 IRS5~}
\newcommand{\verbatimfont}[1]{\def\verbatim@font{#1}}%

\begin{document} 

   \title{FAUST XXIV. Large dust grains in the protostellar outflow cavity walls of the Class I binary L1551 IRS5.}
   \nolinenumbers
   \titlerunning{FAUST XXIV. Large dust grains in the dusty cavity walls of L1551 IRS5}
   \authorrunning{Sabatini et al.}

   \author{G.~Sabatini\inst{1}\orcidlink{0000-0002-6428-9806}
   \and
   {E.~Bianchi}\inst{1}\orcidlink{0000-0001-9249-7082}\and
   {C.~J.~Chandler}\inst{2}\orcidlink{0000-0002-7570-5596}\and
   {L.~Cacciapuoti}\inst{3}\orcidlink{0000-0001-8266-0894}\and
   {L.~Podio}\inst{1}\orcidlink{0000-0003-2733-5372}\and
   {M.~J.~Maureira}\inst{4}\orcidlink{0000-0002-7026-8163}\and
   {C.~Codella}\inst{1,5}\orcidlink{0000-0003-1514-3074}\and
   {C.~Ceccarelli}\inst{5}\orcidlink{0000-0001-9664-6292}\and
   {N.~Sakai}\inst{6}\orcidlink{0000-0002-3297-4497}\and
   {L.~Testi}\inst{7}\orcidlink{0000-0003-1859-3070}\and
   {C.~Toci}\inst{8}\orcidlink{0000-0002-6958-4986}\and
   {B.~Svoboda}\inst{2}\orcidlink{0000-0002-8502-6431}\and
   {T.~Sakai}\inst{9}\orcidlink{0000-0003-4521-7492}\and
   {M.~Bouvier}\inst{10}\orcidlink{0000-0003-0167-0746}\and
   {P.~Caselli}\inst{4}\orcidlink{0000-0003-1481-7911}\and 
   {N.~Cuello}\inst{5}\orcidlink{0000-0003-3713-8073}\and
   {M.~De Simone}\inst{8,1}\orcidlink{0000-0001-5659-0140}\and   
   {I.~J\'{i}menez-Serra}\inst{11}\orcidlink{0000-0003-4493-8714}\and
   {D.~Johnstone}\inst{12,13}\orcidlink{0000-0002-6773-459X}\and   
   {L.~Loinard}\inst{14,15,16}\orcidlink{0000-0002-5635-3345}\and
   {Z.~E.~Zhang}\inst{17}\orcidlink{0000-0002-9927-2705}\and
   {S.~Yamamoto}\inst{18}\orcidlink{0000-0002-9865-0970}
   }
      \institute{INAF, Osservatorio Astrofisico di Arcetri, Largo E. Fermi 5, I-50125, Firenze, Italy; \email{giovanni.sabatini@inaf.it} 
      \and National Radio Astronomy Observatory, PO Box O, Socorro, NM 87801, USA;
      \and European Southern Observatory, Alonso de Cordova 3107, Vitacura, Region Metropolitana de Santiago, Chile;
      \and Center for Astrochemical Studies, Max-Planck-Institut f\"{u}r extraterrestrische Physik, Gie{\ss}enbachstr. 1, 85748 Garching, Germany;
      \and Univ. Grenoble Alpes, CNRS, IPAG, 38000 Grenoble, France;
      \and RIKEN Cluster for Pioneering Research, 2-1, Hirosawa, Wako-shi, Saitama 351-0198, Japan;
      \and Dipartimento di Fisica e Astronomia “Augusto Righi” Viale Berti Pichat 6/2, Bologna, Italy;
      \and European Southern Observatory, Karl-Schwarzschild-Strasse 2, D-85748 Garching bei M\"{u}nchen, Germany;
      \and Graduate School of Informatics and Engineering, The University of Electro-Communications, Chofu, Tokyo 182-8585, Japan;
      \and Leiden Observatory, Leiden University, P.O. Box 9513, 2300 RA Leiden, The Netherlands;
      \and Centro de Astrobiolog\'{i}a (CSIC-INTA), Ctra. de Torrej\'{o}n a Ajalvir, km 4, 28850, Torrej\'{o}n de Ardoz, Spain;
      \and NRC Herzberg Astronomy and Astrophysics, 5071 West Saanich Rd, Victoria, BC, V9E 2E7, Canada;
      \and Department of Physics and Astronomy, University of Victoria, Victoria, BC, V8P 5C2, Canada;
      \and Instituto de Radioastronomía y Astrofísica, UNAM, Apartado Postal 3-72, Morelia 58090, Michoacán, Mexico;
      \and Black Hole Initiative at Harvard University, 20 Garden Street, Cambridge, MA 02138, USA;
      \and David Rockefeller Center for Latin American Studies, Harvard University, 1730 Cambridge Street, Cambridge, MA 02138, USA;
      \and Star and Planet Formation Laboratory, RIKEN Cluster for Pioneering Research, Wako, Saitama 351-0198, Japan;
      \and SOKENDAI (The Graduate University for Advanced Studies), Shonan Village, Hayama, Kanagawa 240-0193, Japan;
      }

   \date{Received:~26~March~2025~|~Accepted:~15~May~2025}
 
  \abstract
   {Planet formation around young stars requires the growth of interstellar dust grains from $\mu$m-sized particles to km-sized planetesimals. Numerical simulations have shown that large ($\sim$mm-sized) grains found in the inner envelope of young protostars could be lifted from the disc via winds. However we are still lacking unambiguous evidence for large grains in protostellar winds/outflows.}
   {We investigate dust continuum emission in the envelope of the Class I binary \IRSFAUST in the Taurus molecular cloud, aiming to identify observational signatures of grain growth, such as variations in the dust emissivity index ($\beta_{\rm mm}$).}%
   {In this context, we present new, high-angular resolution (50 au), observations of thermal dust continuum emission at 1.3~mm and 3~mm in the envelope ($\sim$3000~au) of \IRSFAUST, obtained as part of the ALMA-FAUST Large Program.}
   {We analyse dust emission  along the cavity walls of the CO outflow, extended up to $\sim$1800~au. We find an H$_2$ volume density $>$2$\times$10$^{5}$~cm$^{-3}$, a dust mass of $\sim$58~$M_\Earth$, and $\beta_{\rm mm}\lesssim 1$, implying the presence of grains $\sim$10$^3$ times larger than the typical ISM sizes.}
   {We provide the first spatially resolved observational evidence of large grains within an outflow cavity wall. Our results suggest that these grains have been transported from the inner disc to the envelope by protostellar winds and may subsequently fall back into the outer disc by gravity and/or via accretion streamers. This cycle provides longer time for grains to grow, playing a crucial role in the formation of planetesimals.}
   

   \maketitle
%
   \nolinenumbers

\section{Introduction}\label{sec1:intro}
The formation of terrestrial planets and rocky cores of giant planets is a complex process harboured inside protoplanetary discs surrounding young stellar objects (YSOs). This process begins with submicron-sized dust grains -- typically from 0.01 to 0.3~$\mu$m in the diffuse interstellar medium (ISM; \citealt{Mathis77}) and up to 1~$\mu$m in dense molecular clouds (\citealt{Dartois24}) -- and culminates in the disc, with the formation of dust pebbles, planetesimals, and planets that may harbour life \citep[e.g.][]{Testi14, Drazkowska23, Birnstiel24}.\\  
\indent The Atacama Large Millimeter/submillimeter Array (ALMA) has enabled significant progress in understanding the initial conditions of planet formation. ALMA observations have demonstrated that dust grains in protoplanetary discs can reach millimeter/centimeter sizes (e.g. \citealt{Testi14, Liu2021, Macias21, Radley25}), consistent with predictions from dust growth/evolution models \citep[e.g.,][]{Birnstiel12}, while also suggesting that the dust observed in more evolved Class II discs is likely enriched with a secondary generation of grains \citep[e.g.][]{Turrini19, Guidi22, Testi22}.
However, a major hurdle arises when considering the fate of larger ($\gtrsim$ mm-sized) grains in protoplanetary discs. Models predict that the dust experiences drag forces due to the different velocities between the dust grains and the gas in the disc, causing the mm/cm-size dust grain to progressively spiral inwards towards the central YSO, with the fastest velocity associated with the grains marginally coupled to the gas. These grains have Stokes numbers close to unity and their sizes range between 1~mm and 1~m for typical disc parameters \cite[e.g.][]{Weidenschilling77, Laibe12}. This migration is also supported by observations, showing that large grains are concentrated toward the centre of Class II discs (e.g. \citealt{CarrascoGonzalez19, Tazzari21}). The so-called ``radial drift'' of large grains occurs on timescales much shorter \citep[e.g.][]{Laibe12} than the disc lifetime (a few Myr; \citealt{Ribas15}) posing a major challenge to the theory of planet formation.\\
\indent More recent observations suggest that dust growth up to mm-size could begin in collapsing inner protostellar envelopes ($\sim$500~au), much earlier and further away from the host stars than previously thought \citep[e.g.][]{Miotello14, Galametz19, Bouvier21}. 
Despite numerical simulations predict inefficient dust growth at the typical densities of collapsing protostellar envelopes ($\sim$10$^5$~cm$^{-3}$; e.g. \citealt{Guillet20, Bate22}), other studies suggest that large grains from the inner disc, where the dust growth is efficient due to the large density, could be transported into the envelope via magnetohydrodynamic (MHD) winds \citep[e.g.][]{Giacalone19, Tsukamoto21}. This scenario is supported by the anti-correlation between the dust spectral index and the jet mass-loss rate observed in Class 0 protostars \citep{Cacciapuoti24}. Previous results from the ALMA Large Program (LP) ``Fifty AU STudy of the chemistry in the disc/envelope system of Solar-like protostars'' (FAUST)\footnote{\url{http://faust-alma.riken.jp}; see \citet{Codella21}} suggested the presence of dust-rich cavity walls associated with the outflow driven by the Class~0/I~IRS7B system in the Corona Australis cluster \citep{Sabatini24}. However, the detection of emission in these structures only at $\sim$1.3~mm hampered the estimate of the dust emissivity index ($\beta_{\rm mm}$; that 
anti-correlates with the max grain size) to test the scenario of dust wind transport.\\
\indent This Letter presents the first direct evidence of large dust grains in the cavity wall of spatially resolved protostellar outflow, based on new ALMA continuum observations at 1.3 and 3~mm towards L1551 IRS5 obtained as part of FAUST LP.\\
\indent L1551 IRS5 is a Class I binary system in the Taurus star-forming region. The two protostars, separated by 50~au,have a total mass of 1~$\Msun$ \citep{Rodriguez98, Hernandez24}. Both protostars drive jets \citep{Rodriguez03jets, Feeney23} and harbour hot corino chemistry \citep{Bianchi20, CruzSaenzdeMiera25}. The binaries are surrounded by a circumbinary disc with a radius of $\sim$140~au and a total mass of $\sim$0.02-0.03~$\Msun$ \citep[e.g.][]{Takakuwa20, Kospal21}. The source has been classified as an FU Orionis-like object \citep{Connelley18} and has a total bolometric luminosity $L_{\rm bol}$$\sim$27~L$_\odot$ (for a distance of 147~pc; \citealt{Green13}). The structure and kinematics of the large-scale envelope has been extensively investigated \citep[e.g.][]{Osorio03}.
This makes L1551 IRS5 an excellent candidate to investigate the dust growth process from small to envelope scales, exploiting multi-scale and multi-wavelengths high-sensitivity observations of ALMA-FAUST.

\section{Observations and data reduction}\label{sec2:data}
We observed the \IRSFAUST system with ALMA between October 2018 and August 2019 as part of the FAUST LP (2018.1.01205.L; PI: S. Yamamoto). These observations achieved a native angular resolution of 0$\farcs$4$\times$0$\farcs$3; $\sim$52 au at the source distance of 147$\pm$5 pc (\citealt{Connelley18}), and a maximum recoverable scale of $\sim$21$\arcsec$~($\sim$3100~au). In this Letter, we present data obtained in two spectral setups: Band 3 (85-101 GHz) and Band 6 (214-234 GHz), using both the 12-m array and the Atacama Compact Array (ACA; Band 6 only; see App.~\ref{App:data}).\\
\indent All observations were made with a water vapour $\lesssim$~3.0~mm and centred on $\alpha_{\rm J2000}= {\rm 04^{h}31^{m}34.^{s}14}$, $\delta_{\rm J2000} = {\rm +18^\circ08\arcmin05\farcs10}$. Data calibration was performed using a modified version of the CASA pipeline (v5.6.1-8; \citealt{CASA_Team_22}), including additional routines to correct for $T_{\rm sys}$ and spectral normalisation issues\footnote{\url{https://help.almascience.org/index.php?/Knowledgebase/Article/View/419}}. The data were self-calibrated using line-free channels, improving the dynamic range of the maps by a factor $\sim$5. The final continuum maps were cleaned using {\verb~deconvolver~} =`mtmfs'~({\verb~nterms~}~=~2) and natural weighting, a common uv-range in both bands (inner uv-cutoff: {\verb~uvrange~} =`${\rm >10klambda}$') and a restoring beam of 0\farcs5. The typical rms noise in the maps is $\sim$0.05~mJy~beam$^{-1}$ (B6; 1.3~mm) and $\sim$0.04~mJy~beam$^{-1}$ (B3; 3.0~mm). The absolute flux uncertainty is found to be $<$10\%. The primary beam (PB) correction was applied to all the images.

\begin{figure*}
   \centering
   \includegraphics[width=1\hsize]{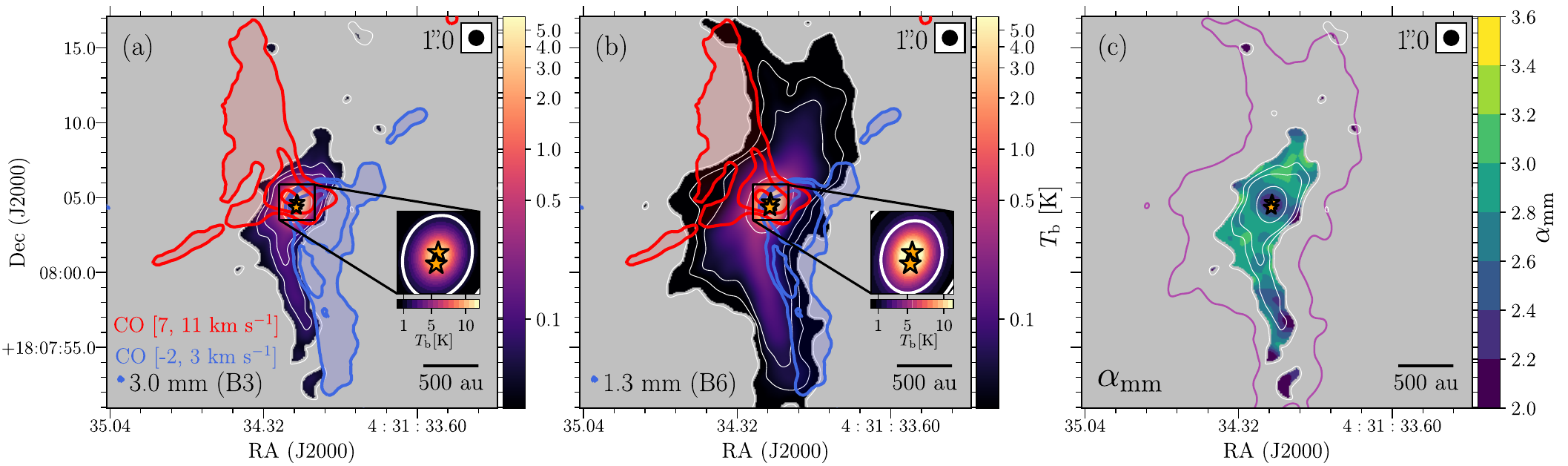}
   \caption{The L1551 IRS5 Class I binary system. $(a,b)$ FAUST maps at 1.3 and 3.0~mm, at the resolution of 1\farcs0. White contours show the [3, 6, 10, 100]$\sigma$, with $\sigma_{\rm 1.3~mm}= 0.13$~mJy~beam$^{-1}$ and $\sigma_{\rm 3.0~mm}= 0.08$~mJy~beam$^{-1}$. The red- and blue-shifted CO~(2-1) emission show the [3, 12, 33]$\sigma_{\rm CO}$ emission ($\sigma_{\rm CO}= 0.5$~Jy~beam$^{-1}$~km~s$^{-1}$). $(c)$ Dust spectral index map. White contours follow those in (a), while the purple one shows the 3$\sigma$ level in (b). Maps are masked at 3$\sigma$ and the beams are shown in the upper corners.} 
\label{fig:cont_tau}%
\end{figure*}

\section{Results}\label{sec3:analisys}
Figure~\ref{fig:cont_tau} shows the distribution of the dust continuum emission at 3.0~and 1.3~mm toward L1551 IRS5. To improve the signal-to-noise ratio, both maps have been smoothed to 1\farcs0~angular resolution. The two circumstellar discs, L1551 IRS5-N and IRS5-S, and their circumbinary disc are shown at the center of the field. \\
\indent The ALMA continuum emission at both wavelengths also reveals an elongated structure extending towards the South. This structure is detected above the 6$\sigma$ level up to $\sim$1300 au (in B3) and $\sim$1800 au (in B6) from L1551 IRS5. A second, fainter and less extended counterpart is detected toward the North, visible only at 1.3 mm. Comparison with red- and blue-shifted CO (2--1) emission (2016.1.00209.S; \citealt{Cruz-SaenzDeMiera19}) indicates that these continuum structures trace the edges of the cavities opened by the blue- (southwest) and red-shifted (northeast) lobes of the outflow driven by L1551~IRS5, with a wide opening angle of $\sim$90$^\circ$ \citep[e.g.][]{Wu09}. 

\subsection{Physical properties of the dusty cavity walls}\label{sec3.1:PhysicalProp}
To calculate the mass budget and H$_2$ column density associated with the observed dusty cavity walls, we assume that the thermal balance of the \IRSFAUST envelope is dominated by central heating from the accreting protostars. We can therefore derive a map of dust temperature ($T_{\rm dust}$) following \cite{Motte01} and \cite{Jorgensen02}, where $T_{\rm dust}(r) = {\rm 38~K}\times(L_{\rm bol}/\Lsun)^{0.2}\times(\it r/ \rm 100~{\rm au})^{-0.4}$. The dust temperature along the cavity wall ranges from $\sim$25 to $\sim$50~K. The continuum opacity can then be computed as $\tau = - {\rm ln}[1 - I_{\nu}/B_\nu$($T_{\rm dust})]$, where $I_\nu$ is the observed specific intensity and $B_\nu$($T_{\rm dust})$ is the Planck function at $T_{\rm dust}$. The $\tau$ derived at 1.3 and 3.0~mm are always $\lesssim$0.20, implying optically thin emission (Fig. \ref{fig:taumaps}).\\
\indent To derive the mass of dust along the cavity walls, $M_{\rm dust}$, and the gas column density, $N$(H$_2$), we assumed $T_{\rm dust} = 30$~K, corresponding to the median value along the cavity walls, and a dust opacity at 1.3~mm, $\kappa_{\rm 1.3mm}$=~0.9 cm$^2$ g$^{-1}$, appropriate for icy-mantled dust grains at a density of $\sim$10$^6$ cm$^{-3}$ \citep{Ossenkopf94}, which is a reasonable assumption in the cold dusty cavities. From an integrated intensity of 55~mJy along the southern cavity wall, we infer a mass of dust in the cavity of $M_{\rm dust}$~$\sim$~58~M$_{\Earth}$. This corresponds to $\sim$40$\%$ of the total dust mass in the \IRSFAUST system, i.e. in the circumstellar discs (N and S), and the circumbinary disc surrounding them \citep[$\sim$140~M$_{\Earth}$, ][]{Cruz-SaenzDeMiera19}.  Assuming a dust-to-gas ratio of 100, we infer a total $N$(H$_2$)~$\sim$3.6$\times$10$^{21}$~cm$^{-2}$ over the structure, reaching its peak of $\sim$7.1$\times$10$^{20}$~cm$^{-2}$ at the location corresponding to the maximum intensity along the cavity wall ($I^{\rm max}_{\rm 1.3mm}$~$\sim$~$11$~mJy beam$^{-1}$). Assuming that the length of the cavity wall along the line of sight is equal to its width in B3, $\sim$250~au, we estimate a lower limit of the average H$_2$ volume density along the cavity walls of $n$(H$_2$)~$>$~2$\times$10$^{5}$~cm$^{-3}$. Finally, we note that the temperature in the outflow cavities may be a factor $1.2-1.5$ higher than what estimated in this section, if there is a additional heating due to UV irradiation and/or slow shocks \citep[e.g.][]{Visser12, Lee15}. Considering this temperature variation, and assuming an intrinsic uncertainty of $\sim$20\% for the assumed gas-to-dust ratio \citep{Sanhueza17}, depending on grain size, shape, and composition along the cavity wall, the resulting masses and densities may be reduced by $\sim$25-40\% compared to those estimated assuming 30~K.

\subsection{Dust spectral index in the dusty cavity walls}\label{sec3.2:alphaindex}
To constrain the maximum size of the grains emitting along the cavity walls, we use the maps of continuum emission in ALMA Band 3 and 6 (see Fig.~\ref{fig:cont_tau}) to obtain a map of the slope ($\alpha_{\rm mm}$) of the mm spectral energy distribution (SED; $F_\nu \propto \nu^{\alpha_{\rm mm}}$): 
\begin{equation}\label{eq:alphaindex}
    \alpha_{\rm mm} = \frac{{\rm log_{10}}F(\nu_{\rm B3}) - {\rm log_{10}}F(\nu_{\rm B6})}{{\rm log_{10}}(\nu_{\rm B3}) - {\rm log_{10}}(\nu_{\rm B6})},
\end{equation}
\noindent where $F(\nu_{\rm B3})$ and $F(\nu_{\rm B6})$ are the PB-corrected fluxes in B3 and B6, respectively, while $\nu_{\rm B3} = 100.6$~GHz and $\nu_{\rm B6} = 225.4$~GHz are the representative frequencies of the two ALMA Bands. The PB correction (Sect.~\ref{sec2:data}) along the cavity walls leads to an increase in the observed flux by a factor of $<$1.3 in B6 and $<$1.1 in B3.\\
\indent Figure \ref{fig:cont_tau} shows the variation of $\alpha_{\rm mm}$ where emission at both bands is detected with a signal-to-noise ratio $>$3, corresponding to the region delimitated by the white contours. The corresponding absolute error, $\delta(\alpha_{\rm mm})$, was obtained via standard error propagation including the system calibration uncertainties (5\% in B3 and 10\% in B6; \citealt{Cortes22}) and the uncertainties on the observed ALMA fluxes (see Fig.~\ref{fig:alphaerrmap}). We find a mean $\langle\alpha_{\rm mm}\rangle = 2.68\pm0.26$ in the cavity wall, with values ranging from $2.06$ to $3.04$ (corresponding to the 5th and 95th percentiles of the $\alpha_{\rm mm}$ distribution). These values are consistent across the entire structure, and no clear variation is observed for $\alpha_{\rm mm}$ at different positions along the cavity wall. It is worth noting that thermal free-free contamination is not expected to be relevant at mm wavelengths. To the best of our knowledge, the only available estimate of free-free contamination in a cavity wall is for a distance of $\sim$400~au from the source Serpens SMM1. This intermediate-mass Class 0 protostar is associated with a powerful free-free radio jet. The estimate indicates a free-free contribution of $\sim$15\% at 7~mm in the cavity, which would decrease to less than 1\% at 3~mm \citep{Hull16}. Nevertheless, future analysis based on cm-wavelengths will provide direct constrains on the fraction of free-free contamination in the L1551 IRS5 cavity walls.\\
\indent In Sect.~\ref{sec3.1:PhysicalProp} we find that dust emission is optically thin at both wavelengths and that the Rayleigh-Jeans approximation ($h\nu/k_{\rm B}T\ll 1$) is always valid for the derived $T_{\rm dust}$, which is in the range 25--50 K. 
Therefore, the slope of the dust opacity power-law $\kappa_\nu~\propto~\nu^{\beta_{\rm mm}}$, does not depend on $T_{\rm dust}$ and can be derived as $\beta_{\rm mm} = \alpha_{\rm mm} -2$, where $\beta_{\rm mm}$ is connected with the maximum grain sizes of the dust population \citep{Draine06, Ricci10, Testi14}. To investigate the variation of grain properties along the cavity walls, we evaluate the dust temperature, brightness temperature ($T_{\rm b}$), continuum opacity, $\alpha_{\rm mm}$ and $\beta_{\rm mm}$ in circular regions, with equivalent area of three resolution beams, at specific locations along the blue- and red-shifted cavities. These positions correspond to the blue and red crosses in Figure~\ref{fig:taumaps}, respectively. The obtained profiles of $T_{\rm dust}$, $T_{\rm b}$, $\tau$, $\alpha_{\rm mm}$ and $\beta_{\rm mm}$ are shown in Appendix~\ref{App:radial_profs} and Figure~\ref{fig:profiles}, where the lower-limits on $\alpha_{\rm mm}$ and $\beta_{\rm mm}$ are estimated assuming a 3$\sigma$ upper-limit at~3.0~mm. The bottom panels of Figure~\ref{fig:profiles} show that along the cavity walls $\beta_{\rm mm}\lesssim1$, well below the typical value of $\beta_{\rm ISM} \sim 1.6$ (\citealt{Schwartz82} and \citealt{PlanckColl14}). Regardless of the chemical composition, porosity, and icy-coverage of the grains, values of $\beta_{\rm mm}$~<~1 (for optically thin emission) indicate grains larger than those typically observed in the ISM \citep{Testi14, Ysard19}.

\section{Discussion}\label{sec4:discussion}
The dust emissivity index is sensitive to dust composition, porosity, and the presence/absence of icy mantles \citep{Testi14}. While variations in $\beta_{\rm mm}$ reflect differing grain properties, values below $\sim$1, such as those derived in the previous section, provide strong evidence for the presence of (sub)mm-sized grains within the observed distribution \citep{Draine06, Kholer15, Ysard19}.  Although further laboratory studies are required to better constrain the relationship between $\beta_{\rm mm}$ and dust properties, the presence of grains significantly larger than typical ISM sizes remains the most compelling explanation for our results. This result may provide an additional explanation for the low $\beta_{\rm mm}$ derived at envelope scales towards some Class~0/I sources \citep[e.g.][]{Kwon09, Miotello14, Chen16, Galametz19, Bouvier21}. In this context, we consider two scenarios: ($i$) the dust growth/coagulation occurs in-situ, within the outflow cavity walls, producing, on envelope scales, grains larger than those typically observed in the ISM, and ($ii$) large grains grow in the protostellar disc and are entrained into the envelope by outflows/winds.\\
\indent The first hypothesis could be explained by the compression of envelope material and the resulting local increase in gas and dust density at the edge of the cavity. Simulations of dust evolution in collapsing protostellar envelopes have shown that, at the typical densities of these environments ($\sim$10$^5$~cm$^{-3}$), grains can grow only up to a few microns (e.g. \citealt{HirashitaOmukai09, Bate22} and \citealt{Lebreuilly23}). \cite{Silsbee22} recently showed that, starting with an initial grain size of 1 $\mu$m, only when local gas densities are $>$10$^7$ cm$^{-3}$, the growth timescale required to form mm-sized grains becomes shorter than the estimated lifetime of \IRSFAUST ($\sim$10$^{5}$~yr; \citealt{Fridlund02}). 
Therefore, local in-situ grain growth may potentially play a role  only in specific over-densities along the walls, where $n$(H$_2$) is about two orders of magnitude higher than the lower-limit, $n$(H$_2$)~$>$~2$\times$~10$^5$~cm$^{-3}$, estimated in Sect.~\ref{sec3.1:PhysicalProp}.\\
\indent The second possibility is to consider that the grains are grown in the protostellar discs and are subsequently entrained into the envelope by outflows/winds. \citet{Duchene2024} found evidence for grains up to $\gtrsim$10 $\mu$m in size in the upper layers of the protoplanetary disc around the young star J04202144+2813491 using observations from the James Webb Space Telescope (JWST). These grains are significantly larger (3 to 10$^3$ times) than typical maximum grain sizes ($a_{\rm max}$) in the diffuse ISM and in dense molecular clouds (e.g. \citealt{Weingartner01}). They proposed that these particles, fully coupled to the gas, are entrained into the upper layers by photoevaporating disc-winds. However, mm-sized grains are expected to be moderately settled in the mid-plane of Class~I discs (\citealt{Villenave2023}). Nevertheless, laboratory experiments have shown that the sticking properties of dry dust grains increase by a factor of $\sim$100 when the temperature reaches $\sim$1000-1300~K \citep{Bogdan2020,Pillich2021}. This strongly supports the potential presence of large grains in the inner disc regions, close to the launching point of the winds. Models of dust growth and evolution predict a radial distribution of grain sizes, with the largest grains located closer to the protostar \citep[e.g.][]{Garaud07, Birnstiel10}. Notably, the presence of dust particles with sizes larger than $\gtrsim$1~mm has been inferred in the inner $\sim$10-20~au of Class II protoplanetary discs \citep[e.g.][]{Perez12, Perez15, Sierra21} and in the source FU Ori \citep[e.g.][]{Liu2021}.\\ 
\indent Recent solutions of MHD winds models have shown that grains can be lifted from the disc to the envelope \citep[e.g.][]{Giacalone19, Tsukamoto21, Bhandare2024}. In particular, \cite{Giacalone19} derive the $a_{\rm max}$ which can be entrained into the envelope following the simplified equation\footnote{Eq.~\ref{eq:amax_Giacal19} assumes an outflow's launching footpoint $r = 1$~au, a disc ﬂaring ratio $z/r = 0.1$, and a ratio between disc's outer and inner edge ${r_{+}/r_{-}} = 500$ (i.e. typical disc of 50~au and inner edge of 0.1~au). 
We refer to \cite{Giacalone19} for more details.}:
\begin{equation}
   a_{\rm max}~\propto~0.24~\Bigg(\frac{\dot M}{10^{-8}~{\rm M_{\odot}/yr}} \Bigg)~\Bigg(\frac{T_{\rm gas}}{200~{\rm K}} \Bigg)^{0.5}\Bigg(\frac{M_*}{{\rm M_{\odot}}}\Bigg)^{-1}~\mu m.    \label{eq:amax_Giacal19}
\end{equation}
Based on Eq.~\ref{eq:amax_Giacal19}, $a_{\rm max}$ is proportional to the mass-loss rate, $\dot{M}$, the gas temperature, $T_{\rm gas}$, and the stellar mass, $M_*$. For L1551 IRS5, we assumed $M_* = 1$~M$_\odot$ (Sect.~\ref{sec1:intro}), $T_{\rm gas} = 100$~K as fiducial value for the gas temperature in protostellar jets \citep[e.g.][]{Podio21}, $\dot M\sim$(1--5)$\times$10$^{-5}$~M$_\odot$~yr$^{-1}$ estimated from molecular emission \citep{Fridlund02}, and we obtain $a_{\rm max}$$\sim$[0.2 -- 0.9]~mm, depending on the assumed $\dot M$. These values are up to three orders of magnitude larger than the typical ISM values, and align with the calculated $\beta_{\rm mm}$$\sim$0.7 (Sect.~\ref{sec3.2:alphaindex}; e.g. \citealt{Testi14, Ysard19}). Furthermore, L1551 IRS5, classified as a FU Ori object, may have experienced outbursts with significantly larger accretion and mass-loss rates. Assuming the grains survive the intense radiation field without sublimating \citep{ChauGiang25}, at these high $\dot{M}$ even large grains (sizes $>$1~mm) at the base of the outflow ($<$1~au; \citealt{Bourdarot2023}) can be entrained by the winds and populate the observed dusty cavity walls. This scenario is supported by the anti-correlation between $\beta_{\rm mm}$ and the jet mass-loss rate for the CALYPSO sample of Class 0 protostars \citep[e.g.][]{Galametz19, Cacciapuoti24}. In this context, higher-angular resolution observations would be crucial to distinguish between these two scenarios. If large grains are transported from the inner disc by MHD winds, we expect lower $\beta_{\rm mm}$ values near the wind/jet launching point, where large grains are entrained from the disc. Moving outward from the protostar, $\beta_{\rm mm}$ should increase, eventually reaching the typical ISM value. Conversely, if in-situ dust growth/coagulation occurs along the dusty cavity walls, we expect lower $\beta_{\rm mm}$ values at the positions of local overdensities along the walls.\\
\indent Regardless the physical mechanism behind the observed low $\beta_{\rm mm}$, we can speculate that large grains along the cavity walls may subsequently fall back onto the protostellar disc due to gravity as predicted by \cite{Tsukamoto21}, allowing grains to further grow up to the necessary limit to overcome the radial drift barrier. This scenario also provide a possible explanation for the observed isotopic anomalies in Solar System meteorites, i.e. the presence of refractory inclusions in chondrites, such as calcium-aluminum-rich inclusions (CAIs) and amoeboid olivine aggregates (AOAs). These inclusions are considered key witnesses to the primordial dust present in the solar protoplenetary disc, representing high-temperature condensates formed near the protosun before being transported to the outer disc \citep[e.g.][]{Morbidelli24}.

\section{Conclusions}\label{sec5:conclusion}
The analysis of dust continuum emission at 1.3 and 3 mm, obtained in the context of ALMA-FAUST LP, allows us to derive the spatially resolved dust spectral index in the cavity walls of the outflow associated with the \IRSFAUST binary system. The low inferred value ($\beta_{\rm mm} < 1$) strongly suggests, for the first time, the presence of mm-sized dust grains along the protostellar outflow cavities.\\
\indent This new result may explain previous evidence of large grains  at envelope scales in Class 0/I sources \citep[e.g.][]{Kwon09,Miotello14,Galametz19}. More specifically, grains can grow in protostellar discs and then be lifted in the envelope by outflow/winds \citep[e.g.][]{Giacalone19, Bhandare2024, Cacciapuoti24}. In light of these results, we also speculate that large grains may successively fall back in the outer disc due to gravity \citep{Tsukamoto21} and/or transportation along accretion streamers \citep{Cacciapuoti24b}. This will provide additional time for grains to grow overcoming the long standing problem of pebble formation in discs \citep[e.g.][]{Drazkowska23, Birnstiel24}.\\
\indent This study underscores the importance of complementing disc observations with high-sensitivity intermediate- to large-scale observations in order to investigate the dust growth process in young protostellar environments. It appears now crucial to expand our pilot study to a larger sample of sources associated with powerful outflows to detect dusty cavity walls, and quantify the role of outflows and winds in the grain growth process. Additional evidence may also come from polarisation studies at comparable spatial scales \citep[e.g.][]{Hull20}.\\
\indent Future astronomical facilities, such as the next-generation Very Large Array (ngVLA)\footnote{Official ngVLA webpage: \url{https://ngvla.nrao.edu/}} and the Square Kilometre Array Observatory (SKAO)\footnote{Official SKAO webpage: \url{https://www.skao.int/en}}, may represent a significant step forward in characterising dusty cavity walls. These facilities will extend high sensitivity and angular resolution observations to radio wavelengths, providing crucial constraints on the physics of dust growth during stars and planets formation process.

\vspace{-5pt}
\begin{acknowledgements}
The authors thank the anonymous referee for her/his suggestions to improve the manuscript. This Letter makes use of the following ALMA data: ADS/JAO.ALMA\#2018.1.01205.L (PI: S. Yamamoto). GS, LP and CC acknowledge the project PRIN-MUR 2020 MUR BEYOND-2p (Prot. 2020AFB3FX), the PRIN MUR 2022 FOSSILS (Chemical origins: linking the fossil composition of the Solar System with the chemistry of protoplanetary disks, Prot. 2022JC2Y93), the project ASI-Astrobiologia 2023 MIGLIORA (F83C23000800005), and the INAF-GO 2023 fundings PROTO-SKA (C13C23000770005). GS thanks the INAF-Minigrant 2023 TRIESTE (PI: G. Sabatini). NC acknowledges the financial support from the ERC under the European Union Horizon Europe programme (g.a. No. 101042275; Stellar-MADE). LP acknowledges the INAF Mini-Grant 2022 “Chemical Origins” (PI: L. Podio). EB is supported by the Next Generation EU funds within the National Recovery and Resilience Plan (PNRR), Mission 4 - Education and Research, Component 2 - From Research to Business (M4C2), Investment Line 3.1 - Strengthening and creation of Research Infrastructures, Project IR0000034 – “STILES - Strengthening the Italian Leadership in ELT and SKA”. MB acknowledges support from the ERC Advanced Grant MOPPEX 833460. LL acknowledges the support of DGAPA PAPIIT grants IN108324 and IN112820 and CONACyT-CF grant 263356. IJ-S acknowledges funding from grants No. PID2019-105552RB-C41 and PID2022-136814NB-I00 from the Spanish Ministry of Science and Innovation/State Agency of Research MCIN/AEI/10.13039/501100011033 and by “ERDF A way of making Europe”. DJ is supported by NRC Canada and by an NSERC Discovery Grant. LL acknowledges the support of UNAM-DGAPA PAPIIT grants IN108324 and IN112820 and CONACYT-CF grant 263356.
\end{acknowledgements}

%
%

\vspace{-20pt}
\bibliographystyle{aa} 
\bibliography{mybib_GAL}

\begin{appendix}
\section{Data reduction details}\label{App:data}

Table \ref{tab:obs_ALMA} summarises the ALMA\footnote{ALMA is a partnership of the ESO (representing its member states), the NSF (USA) NINS (Japan), the NRC (Canada), the NSC and ASIAA (Taiwan), in cooperation with the Republic of Chile. The JAO is operated by the ESO, the AUI/NRAO, and the NAOJ.} setup 1 and setup 3 observations of the \IRSFAUST system conducted as part of the FAUST Large Program (2018.1.01205.L PI: S. Yamamoto; \citealt{Codella21}). The FAUST setup 1 (Band 6) data were acquired using 12 spectral windows (SPWs), each with a bandwidth ($\Delta\nu$) of 58.6 MHz and a frequency resolution ($\delta\nu$) of 122 kHz. In setup 3 (Band 3), 6 SPWs were observed with $\Delta\nu = 58.6$~MHz and a $\delta\nu = 61$~kHz. In all spectral setups, an additional SPW with a bandwidth of 1.875 GHz was dedicated to observing the thermal dust continuum emission ($\delta\nu = 0.49$ MHz, i.e., a velocity resolution of $\sim$0.65~km~s$^{-1}$ in Band 6 and $\sim$1.5~km~s$^{-1}$ in Band 3).\\
\indent Many of the FAUST targets had very complex spectra, including \IRSFAUST, so line-free channels had to be identified by hand (including in the ``continuum'' SPW). These were then used for creating a continuum image and model from which self-calibration gain solutions were derived and applied to all the channels, for each ALMA configuration. A similar self-calibration technique was used when combining data from multiple configurations, which had the effect of correcting for any position offsets, and aligning the amplitude calibration. Continuum images were made for the analysis described in this paper as specified in Section~\ref{sec2:data}.

\begin{table}
        \caption{Technical details of the FAUST observations through the \IRSFAUST system.}\label{tab:obs_ALMA}
	\setlength{\tabcolsep}{10pt}
	\renewcommand{\arraystretch}{1.05}
	\centering
	\begin{tabular}{l|cc}
		\toprule
            Fields & Band~3 & Band~6 \\
		\midrule
$\nu$~range                              &  93--108 GHz       & 216--234 GHz          \\
Configurations                           &  C43-3, C43-6      & C43-2, C43-5, 7M     \\
B$_{\rm min}$-B$_{\rm max}$              &  15--2500~m        & 9--1400m              \\
Antennas$^{(a)}$                         &  46                & 47--51(11--12)        \\
\multirow{2}{*}{$\Delta\nu^{(b)}$}       & 58.6~MHz           & 58.6~MHz              \\
                                         & (175~km~s$^{-1}$)  & (78~km~s$^{-1}$)      \\
\multirow{2}{*}{$\delta\nu^{(b)}$}       & 61~kHz             & 122~kHz               \\ 
                                         & (0.18~km~s$^{-1}$) & (0.16~km~s$^{-1}$)    \\ 
$\langle\theta_{\rm res}\rangle$$^{(c)}$ & $0\farcs5$         & $0\farcs5$            \\
$\langle\theta_{\rm MRS}\rangle$$^{(c)}$ &  10$\arcsec$       & 10$\arcsec$           \\
\multirow{2}{*}{Calibrators}             &  \multicolumn{2}{c}{J0510+1800, J0440+1437}\\
                                         &  \multicolumn{2}{c}{and J0423-0120}        \\
		\bottomrule	
	\end{tabular}
	\tablefoot{$^{(a)}$ Parenthesis refer to the number of antennas for the ACA. Ranges are given when multiple execution blocks are employed; $^{(b)}$ $\Delta\nu$ and $\delta\nu$ are the typical bandwidth and spectral resolution of SPWs dedicated to line observations; $^{(c)}$ restoring beam and maximum recoverable scale for the 10 k$\lambda$ inner uv-cutoff used, as described in Section~\ref{sec2:data}.}
\end{table}

\section{Additional results}
\begin{figure*}
   \centering
   \includegraphics[width=0.8\hsize]{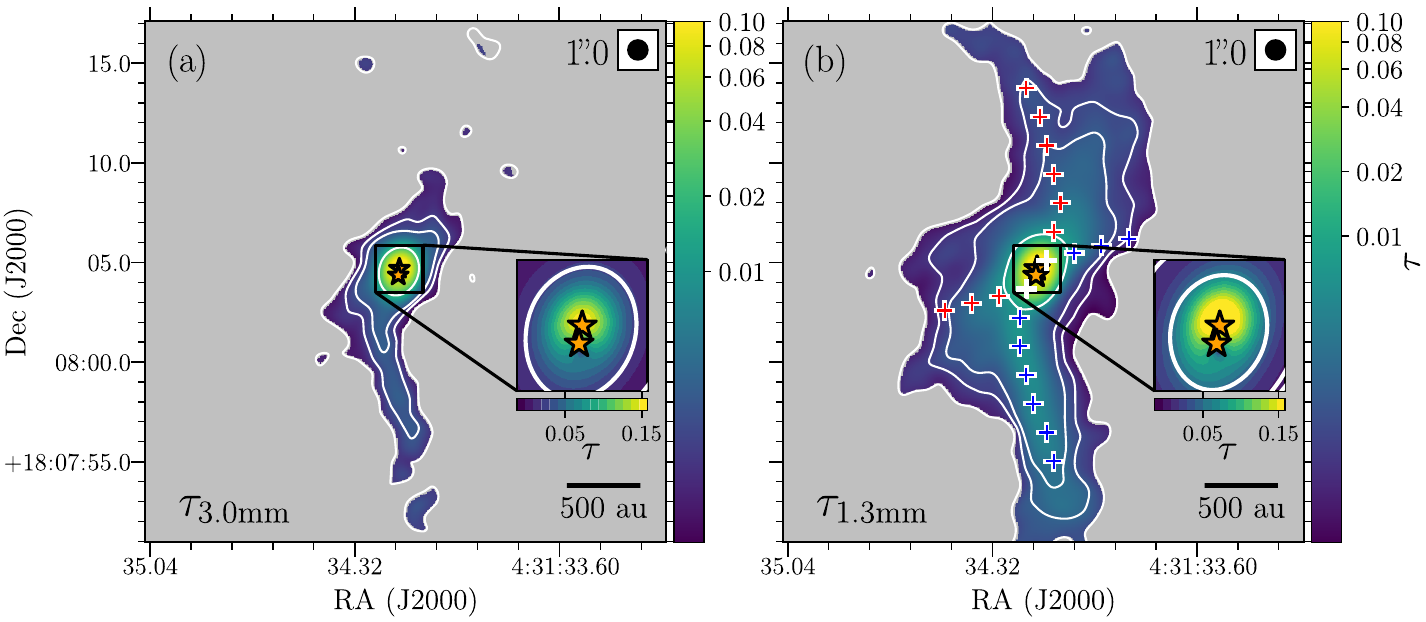}
   \caption{Continuum optical depth derived towards \IRSFAUST at the resolution of 1\farcs0. Panels a and b show the values of $\tau$ derived at 1.3 and 3.0 mm, respectively. White contours follow those in Fig.~\ref{fig:cont_tau}. Crosses indicate the coordinates of the extraction regions associated with the profiles in Fig.~\ref{fig:profiles} (see text Sect.~\ref{sec4:discussion}). Maps are masked at 3$\sigma$ and the beams are shown in the upper corners.}
\label{fig:taumaps}%
\end{figure*}

\begin{figure}
   \centering
   \includegraphics[width=1\hsize]{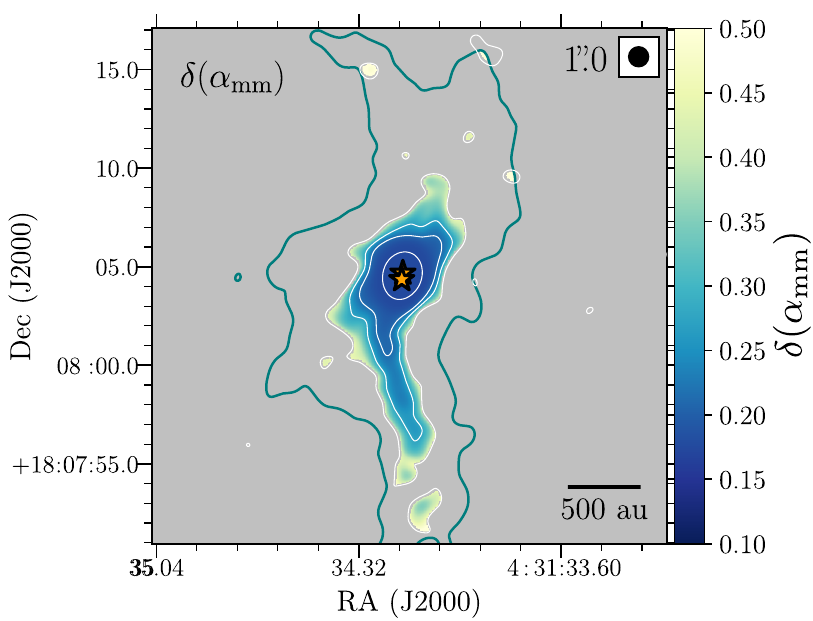}
   \caption{Map of the absolute error, $\delta(\alpha_{\rm mm})$, associated with $\alpha_{\rm mm}$ and derived based on the continuum emission at 1.3 and 3.0~mm as described in Sect. \ref{sec3.2:alphaindex}. White and blue contours are the same as in Fig.~\ref{fig:cont_tau}a.}
\label{fig:alphaerrmap}%
\end{figure}

Figure~\ref{fig:taumaps} shows the continuum opacity derived at 1.3~mm and 3.0~mm, as discussed in Section~\ref{sec3.1:PhysicalProp}. At the center of the field, the opacities obtained in Band 6 are higher than those in Band 3, though both remain below 0.2. This indicates that the recovered emission in both ALMA Bands is optically thin.\\
\indent Figure~\ref{fig:alphaerrmap} displays the absolute error, $\delta(\alpha_{\rm mm})$, associated with the slope of the mm-SED ($\alpha_{\rm mm}$) derived in Section~\ref{sec3.2:alphaindex}. This error map is generated via the standard propagation of the system calibration uncertainties (5\% in Band 3 and 10\% in Band 6; \citealt{Cortes22}) and the uncertainties in the observed ALMA fluxes (i.e., the rms noise in each Band). The absolute $\delta(\alpha_{\rm mm})$ ranges from 0.1 to 0.5, at the edge of the structure, where the signal-to-noise ratio reaches the 3$\sigma$ threshold.

\section{Radial distributions along dusty cavities}\label{App:radial_profs}
\begin{figure*}
\centering
      \includegraphics[width=0.8\hsize]{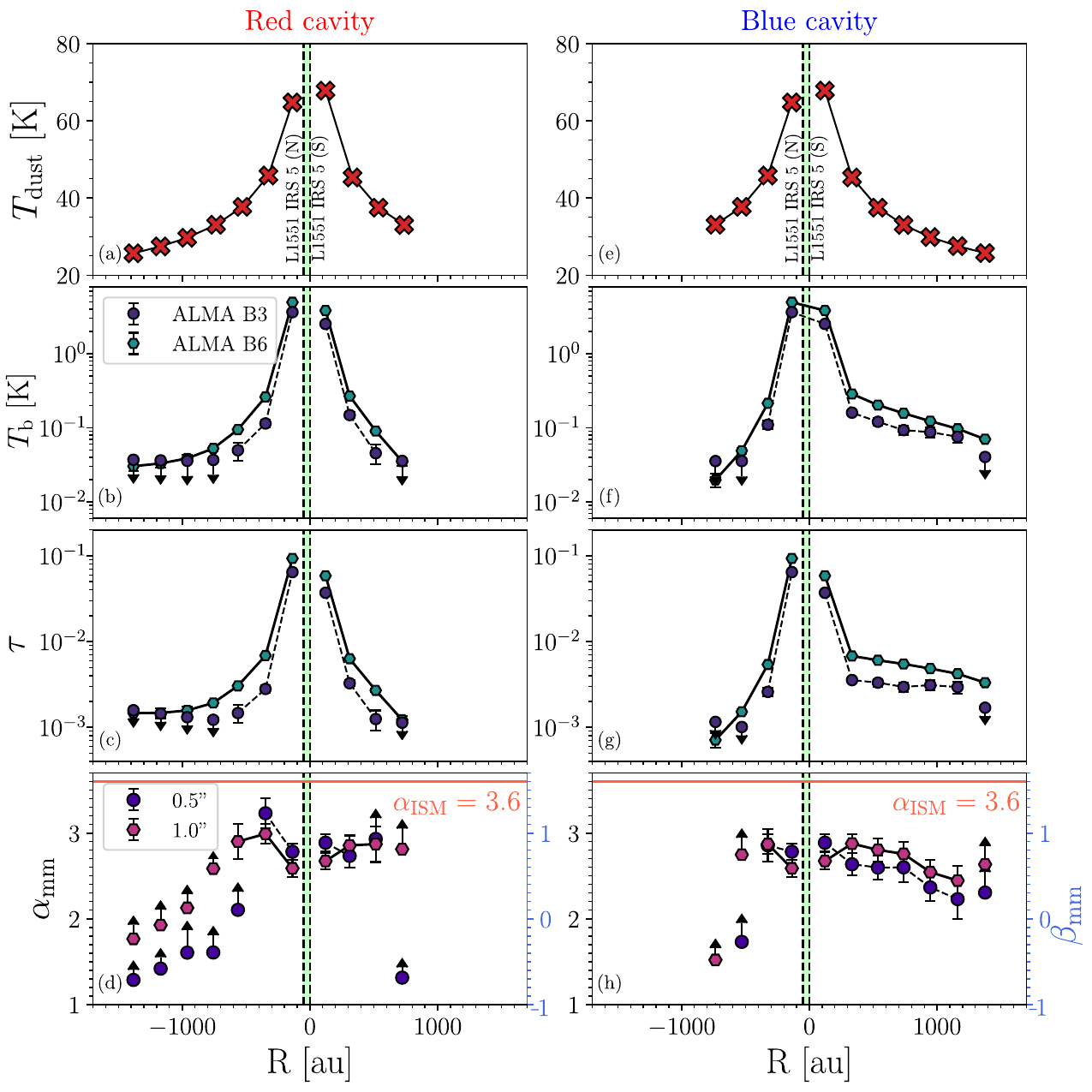}
\caption{Radial profiles of observed and derived physical quantities along the outflow cavities associated with \IRSFAUST - redshifted outflow, left panels (a-d); blueshifted outflow, right panels (e-h). From top to bottom: Dust temperature profiles derived in Sect.~\ref{sec3.1:PhysicalProp} (panels a, e); Brightness temperature observed with ALMA at 1.3 mm (green dots) and 3.0 mm (blue dots; panels b, f) at the resolution of 1\farcs0; Associated optical depth (panels c, g); Dust spectral index ($\alpha_{\rm mm}$) and emissivity index ($\beta_{\rm mm}$) derived at the angular resolution of 0\farcs5 (violet dots) and at 1\farcs0 (pink hexagons; panels d, h).}\label{fig:profiles}%
\end{figure*}

Figure \ref{fig:profiles} summarises the radial profiles of dust temperature, brightness temperature, continuum opacity, dust spectral and emissivity indices along the blue- and redshifted outflow cavity walls of L1551 IRS5. These profiles were derived by averaging values within circular regions, each with an area equivalent to three resolution beams, at the locations indicated by the blue and red crosses in Figure \ref{fig:cont_tau}d. Lower limits on $\alpha_{\rm mm}$ and $\beta_{\rm mm}$ were estimated assuming a 3$\sigma$ upper limit at 3.0~mm where the emission was not detected above this threshold.\\

\end{appendix}
\end{document}